\documentclass[10pt,twocolumn,letterpaper]{article}

\usepackage{cvpr}
\usepackage{times}
\usepackage{epsfig}
\usepackage[usenames, dvipsnames]{color}
\usepackage{xspace}
\usepackage{booktabs}
\usepackage{enumerate}
\usepackage{verbatim}
\usepackage{amsmath}
\usepackage{amssymb}
\usepackage{graphicx}
\usepackage{subcaption}
\usepackage{slashbox}
\usepackage[linesnumbered,ruled]{algorithm2e}
\usepackage{siunitx}
\usepackage{tabu}
\usepackage{soul}
\usepackage[colorlinks=true,breaklinks=true,letterpaper=true,bookmarks=false,urlcolor=black,linkcolor=cyan,citecolor=cyan]{hyperref}
\usepackage{fancyhdr}

\newif\ifauthorcopy
\authorcopytrue

\cvprfinalcopy 
\def\cvprPaperID{10} 

\begin{document}

\newif\ifdraft
\draftfalse

\ifdraft
  \newcommand{\todocolor}[1]{\textcolor{red}{#1}}
  \newcommand{\hlnew}[1]{\hl{#1}}
  \soulregister\cite7
  \soulregister\dnn7
  \soulregister\dnns7
  \soulregister\ref7
  \soulregister\pageref7
  \soulregister\url7
  \soulregister\secref7
  \soulregister\figref7
  \soulregister\emph7
  \soulregister\lpnorm7
\else
  \newcommand{\todocolor}[1]{}
  \newcommand{\hlnew}[1]{#1}
\fi
\newcommand{\mahmood}[1]{\todocolor{[[Mahmood: #1]]}}
\newcommand{\lujo}[1]{\todocolor{[[Lujo: #1]]}}
\newcommand{\mike}[1]{\todocolor{[[Mike: #1]]}}
\newcommand\note[1]{\todocolor{[[Note: #1]]}}
\newcommand\todo[1]{\todocolor{[[#1]]}}

\newcommand\benign[0]{$C_B$}
\newcommand\advimp[0]{$C_\mathit{AI}$}
\newcommand\advper[0]{$C_\mathit{AP}$}

\newcommand{\algref}[1]{\mbox{Alg.~\ref{#1}}}
\newcommand{\secref}[1]{\mbox{Sec.~\ref{#1}}\xspace}
\newcommand{\secsref}[2]{\mbox{Sec.~\ref{#1}--\ref{#2}}\xspace}
\newcommand{\figref}[1]{\mbox{Fig.~\ref{#1}}}
\newcommand{\tabref}[1]{\mbox{Table~\ref{#1}}}
\newcommand{\appref}[1]{\mbox{App.~\ref{#1}}}
\newcommand{\eqnref}[1]{Eqn.~\ref{#1}\xspace}
\newcommand{\eqnsref}[2]{Eqns.~\ref{#1}--\ref{#2}\xspace}

\makeatletter
\DeclareRobustCommand{\varname}[1]{\begingroup\newmcodes@\mathit{#1}\endgroup}
\makeatother

\renewcommand{\ml}[0]{ML}
\newcommand{\lpnorm}[1]{$L_{#1}$}
\newcommand{\frs}[0]{face-rec\-og\-ni\-tion system}
\newcommand{\dnn}[0]{DNN}
\newcommand{\cwloss}[0]{$\Loss_\textit{cw}$}
\newcommand{\subject}[1]{$S_\textit{#1}$}
\newcommand{\dnnFunc}[0]{$F(\cdot)$}
\newcommand{\dnnloss}[0]{$\Loss_\textit{F}$}
\newcommand{\parheading}[1]{\textbf{#1}~\hspace{2pt}}
\newcommand{\smallparheading}[1]{\emph{#1}~\hspace{2pt}}

\newcommand{\gan}[0]{GAN}
\newcommand{\gen}[0]{$G$}
\newcommand{\genloss}[0]{$\Loss_\textit{G}$}
\newcommand{\discrim}[0]{$D$}
\newcommand{\discrimgain}[0]{$\Gain_\textit{D}$}
\newcommand{\latent}[0]{$Z$}

\newcommand{\Loss}[0]{\ensuremath{\mathit{Loss}}}
\newcommand{\Gain}[0]{\ensuremath{\mathit{Gain}}}
\newcommand{\data}[0]{\ensuremath{\mathit{data}}}

\title{On the Suitability of $L_p$-norms for\\
Creating and Preventing Adversarial Examples}

\author{Mahmood Sharif\textsuperscript{\dag}\hspace{8pt}
Lujo Bauer\textsuperscript{\dag}\hspace{8pt}
Michael K. Reiter\textsuperscript{\ddag}\\
\textsuperscript{\dag}Carnegie Mellon University\\
\textsuperscript{\ddag}University of North Carolina at Chapel Hill\\
{\tt\small \{mahmoods,lbauer\}@cmu.edu, reiter@cs.unc.edu}
}

\maketitle

\begin{abstract}

Much research has been devoted to better understanding
adversarial examples, which are specially crafted inputs to
machine-learning models that are perceptually similar to benign inputs, but
are classified differently (i.e., misclassified). Both algorithms that
create adversarial examples and strategies for defending against
adversarial examples
typically use $L_p$-norms to measure the perceptual similarity
between an adversarial input and its benign original. Prior work
has already shown, however, that two images need not be close to each
other as measured by an $L_p$-norm to be perceptually similar. In this
work, we show that nearness according to an $L_p$-norm is not just
unnecessary for perceptual similarity, but is also insufficient.
Specifically, focusing on datasets (CIFAR10 and MNIST),
$L_p$-norms, and thresholds used in prior work, we show
through online user studies that ``adversarial
examples'' that are closer to their benign counterparts than
required by commonly used $L_p$-norm thresholds can nevertheless
be perceptually distinct to humans from the corresponding
benign examples. Namely, the perceptual distance between two
images that are ``near'' each other according to an $L_p$-norm can be
high enough that participants frequently classify the two images as
representing different objects or digits. Combined with prior work,
we thus demonstrate that nearness of inputs as measured by
$L_p$-norms is neither necessary nor sufficient for perceptual
similarity, which has implications for both creating and defending
against adversarial examples.
\hlnew{We propose and discuss alternative similarity metrics to
stimulate future research in the area.}

\end{abstract}

\section{Introduction}


Machine learning is quickly becoming a key aspect of many technologies
that impact us on a daily basis, from automotive driving aids to city
planning, from smartphone cameras to cancer diagnosis. As such, the
research community has invested substantial effort in understanding
\emph{adversarial examples}, which are inputs to machine-learning
systems that are perceptually similar to benign inputs, but that are
misclassified, i.e., classified differently than the benign inputs
from which they are derived (e.g.,~\cite{Biggio13Evasion,Szegedy13NNsProps}).
An attacker who creates an adversarial
example can cause an object-recognition algorithm to incorrectly
identify an object (e.g., as a worm instead of as an panda)~\cite{Gdfllw14ExpAdv},
a street-sign recognition algorithm to fail to recognize a stop
sign~\cite{Evtimov17Signs}, or a face-recognition system to fail to identify a
person~\cite{Sharif16AdvML}. Because of the potential impact on safety
and security, better understanding the susceptibility of machine-learning
algorithms to adversarial examples, and devising defenses, has been a
high priority.

A key property of adversarial examples that makes them dangerous is
that human observers do not recognize them as adversarial. If a human
recognizes an input (e.g., a person wearing a disguise at an airport)
as adversarial, then any potential harm may often be prevented by
traditional methods (e.g., physically detaining the attacker). Hence, most
research on creating adversarial examples (e.g.,~\cite{Carlini17Robustness,
Papernot16Limitations}) or defending against them (e.g.,~\cite{Madry17AdvTraining,
Kolter17Defense}) focuses on adversarial examples that are
\emph{imperceptible}, i.e., a human would consider them perceptually
similar to benign images.

The degree to which an adversarial example is imperceptible from its
benign original is usually measured using $L_p$-norms, e.g., $L_0$
(e.g.,~\cite{Papernot16Limitations}), $L_2$ (e.g.,~\cite{Szegedy13NNsProps}),
or $L_\infty$ (e.g.,~\cite{Gdfllw14ExpAdv}). Informally, for images, $L_0$ measures the
number of pixels that are different between two images, $L_2$ measures the
Euclidean distance between two images, and $L_\infty$ measures the largest difference between
corresponding pixels in two images. These measures of imperceptibility
are critical for creating adversarial examples and defending against them.
On the one hand, algorithms for creating adversarial examples seek to enhance
their imperceptibility by producing inputs that both cause misclassification
and whose distance from their corresponding benign originals has small
$L_p$-norm.
On the other hand, defense mechanisms assume that if the difference
between two inputs is below a specific $L_p$-norm threshold then the
two objects belong to the same class (e.g.,~\cite{Madry17AdvTraining}).

Hence, the completeness and soundness of attacks and defenses commonly rely
on the assumption that some $L_p$-norm is a reasonable measure for
perceptual similarity, i.e., that if the $L_p$-norm of the difference
between two objects is below a threshold then the difference between
those two objects will be imperceptible to a human, and vice
versa. Recent work has shown that one direction of this assumption
does not hold: objects that are indistinguishable to humans (e.g., as a
result of slight rotation or translation) can nevertheless be very
different as measured by the $L_p$-norm of their
difference~\cite{Engstrom17AdvTrans,Kanbak17Transform,Xiao17Transform}.

In this paper we further examine the use of $L_p$-norms as a measure
of perceptual similarity. In particular, we examine whether pairs of
objects whose difference is small according to an $L_p$-norm are
indeed similar to humans. Focusing on datasets, $L_p$-norms,
and thresholds used in prior work, we show that small differences
between images according to an $L_p$-norm do not imply perceptual
indistinguishably. Specifically, using the CIFAR10~\cite{Krizhevsky09CIFAR} and
MNIST~\cite{MNIST} datasets, we show via 
online user studies that images whose distance---as measured by the $L_p$-norm
of their difference---is below thresholds used in prior work can nevertheless
be perceptibly very different to humans. The perceptual distance between
two images can in fact lead people to classify two images differently.
For example, we find that by perturbing about 4\% of pixels in digit
images to achieve small \lpnorm{0} distance from benign images (an
amount comparable to prior work~\cite{Papernot16Limitations}), humans
become likely to classify the resulting images correctly only 3\% of the time.

Combined with previous work, our results show that
nearness between two images according to an $L_p$-norm is
neither necessary nor sufficient for those images to be perceptually
similar. This has implications for both attacks and defenses against
adversarial inputs. For attacks, it suggests that even though a
candidate attack image may be within a small $L_p$ distance
from a benign image, this does not ensure that a human would find
the two images perceptually similar, or even that a human
would classify those two images consistently (e.g., as the same person or
object). For defenses, it implies defense strategies that attempt to
train machine-learning models to correctly classify what ought to
be an adversarial example may be attempting to solve an extremely
difficult problem, and may result in ill-trained machine-learning models.




\hlnew{To stimulate future research on developing better similarity metrics
for comparing adversarial examples with their benign counterparts, we propose
and discuss several alternatives to} \lpnorm{p}-norms.
\hlnew{In doing so, we hope to improve attacks against machine-learning
algorithms, and, in return, defenses against them.}

\hlnew{Next, we review prior work and provide background (\secref{sec:relwork}).
We then discuss the necessity and sufficiency of conditions for perceptual
similarity, and show evidence that \lpnorm{p}-norms lead to conditions that are
neither necessary nor sufficient} (\secsref{sec:limit1}{sec:studies}). \hlnew{Finally, we discuss
alternatives to \lpnorm{p}-norms and conclude} (\secsref{sec:discussion}{sec:conclusion}).

\section{Background and Related Work}
\label{sec:relwork}


In concurrent research efforts, Szegedy et al.\ and Biggio et al.\ showed
that specifically crafted small perturbations of benign inputs can lead
machine-learning models to misclassify them~\cite{Biggio13Evasion,Szegedy13NNsProps}.
The perturbed inputs are referred to as \emph{adversarial examples}~\cite{Szegedy13NNsProps}.
Given a machine-learning model, a sample $\hat{x}$ is considered
as an adversarial example if it is \emph{similar} to a benign sample $x$
(drawn from the data distribution), such that $x$ is correctly classified and
$\hat{x}$ is classified differently than $x$. Formally, for a classification
function $F$, a class $c_x$ of $x$, a distance metric $D$, and a threshold
$\epsilon$, $\hat{x}$ is considered to be an adversarial example if:
\begin{equation}
F(x)=c_x \wedge F(\hat{x})\neq c_x \wedge D(x,\hat{x})\leq \epsilon
\label{eqn:adv_examples}
\end{equation}
The leftmost condition ($F(x)=c_x$) checks that $x$ is correctly classified,
the middle condition ($F(\hat{x})\neq c_x$) ensures that $\hat{x}$ is incorrectly
classified, and the rightmost condition ($D(x,\hat{x})\leq \epsilon$) ensures that 
$x$ and $\hat{x}$ are similar (i.e., their distance is
small)~\cite{Biggio13Evasion}.

Interestingly, the concept of similarity is ambiguous. For example, two images
may be considered similar because both contain the color blue, or because
they are indistinguishable (e.g., when performing ABX tests~\cite{Munson50ABX}).
We believe that prior work on adversarial examples implicitly
assumes that similarity refers to \emph{perceptual or visual similarity},
as stated by Engstrom et al.~\cite{Engstrom17AdvTrans}.
As Goodfellow et al.\ explain, adversarial
examples are ``particularly disappointing because a popular approach in computer
vision is to use convolutional network features as a space where Euclidean
distance approximates perceptual distance''~\cite{Gdfllw14ExpAdv}.
In other words, adversarial examples are particularly of interest
because they counter our expectation that neural networks specifically, and
machine-learning models in general, represent perceptually similar inputs
with features that are similar (i.e., close) in the Euclidean space.

A common approach in prior work has been to use \lpnorm{p}-distance
metrics (as $D$) in attacks that craft adversarial examples
and defenses against them (e.g.,~\cite{Carlini17Robustness}).
For non-negative
values of $p$, the \lpnorm{p} distance between the two $d$-dimensional
inputs $x$ and $\hat{x}$ is defined as~\cite{Riesz10Lp}:
\[||x-\hat{x}||_p =\Big(\sum_{i=1}^{d} |x_i - \hat{x}_i|^p\Big)^{1\over p}\]

The main \lpnorm{p} distances used in the literature are \lpnorm{0},
\lpnorm{2}, and \lpnorm{\infty}. Attacks using
\lpnorm{0} attempt to minimize the number of pixels
perturbed~\cite{Carlini17Robustness,Papernot16Limitations}; those 
using \lpnorm{2} attempt to minimize the Euclidean distance
between $x$ and
$\hat{x}$~\cite{Carlini17Robustness,Szegedy13NNsProps}; and
attacks using \lpnorm{\infty} attempt to minimize the maximum
change applied to any coordinate in $x$~\cite{Carlini17Robustness,Gdfllw14ExpAdv}.

To defend against adversarial examples, prior work has focused on either training
more robust deep neural networks (\dnn{}s) that are not susceptible to small
perturbations~\cite{Gdfllw14ExpAdv,Kantchelian16ICML,Kurakin16AdvTrain,
Madry17AdvTraining,Papernot16Distillation,Szegedy13NNsProps,Kolter17Defense},
or developing techniques to detect adversarial examples~\cite{Feinman17Detector,
Grosse17Detector,Meng17Magnet, Metzen17Detector,Xu17Squeezing}.
\emph{Adversarial training} is a particular defense that has achieved a relatively
high success~\cite{Gdfllw14ExpAdv,Kantchelian16ICML,Kurakin16AdvTrain,
Madry17AdvTraining,Szegedy13NNsProps}.
In this defense, adversarial examples with bounded \lpnorm{p} distance
(usually, $p=\infty$) are generated in each iteration of training \dnn{}s,
and the  \dnn{} is trained to correctly classify those examples. Insufficiency
and lack of necessity in \lpnorm{p}-distance metrics have direct implication on
adversarial training, as we discuss in~\secref{sec:limit1}. 

Despite the goal for adversarial examples to be perceptually
similar to benign samples, little prior work on adversarial examples
has explicitly explored or accounted for human perception.
The theoretical work of Wang et al.\ is an exception, as they treated the
human as an oracle to seek for the conditions under which \dnn{}s would
be robust~\cite{Wang17theoretical}. In contrast, we take
an experimental approach to show that \lpnorm{p} distances may be
inappropriate for defining adversarial examples. \hlnew{Our findings are 
in line with research in psychology, which has found
that distance metrics in geometric spaces may not always match
humans' assessment of similarity (e.g.,~\cite{Tversky77Similarity}). Concurrently
to our work, Elsayed et} al.\ \hlnew{showed that adversarial examples can mislead
humans as well as \dnn{}s~\cite{Elsayed18FoolHuman}. While they considered
images of higher dimensionality than we consider in this work (see
\secref{sec:studies}), they allowed perturbations of higher norm than
commonly found in practice.}

Some work proposed generating adversarial examples via techniques
other than minimizing \lpnorm{p}-norms. In three different concurrent
efforts, researchers proposed to use minimal geometric transformations
to generate adversarial examples~\cite{Engstrom17AdvTrans,
Kanbak17Transform,Xiao17Transform}. As we detail in the next section,
geometric transformations evidence that conditions on \lpnorm{p}-norms
are unnecessary for ensuring similarity. In other work, researchers proposed
to achieve imperceptible adversarial examples by maximizing the Structural
Similarity (SSIM) with respect to benign images~\cite{Riesz10Lp}.
SSIM is a measure of perceived quality of images that has been shown
to align with human assessment~\cite{Wang04SSIM}. It is a differentiable
metric with values in the interval [-1,1] (where a values closer to 1 indicate
higher similarity). By maximizing SSIM, the researchers hoped to increase
the similarity between the adversarial examples and their benign
counterparts. In \secref{sec:discussion} we provide a preliminary analysis
of SSIM as a perceptual similarity metric for adversarial examples.

\hlnew{In certain cases, perceptual similarity to a reference image is
not a goal
for an attack (e.g., images that seem incomprehensible or benign
to humans, but are classified as street signs by machine-vision
algorithms~\cite{Liu18TrojanNN,Nguyen15FoolNN,Sitawarin18Darts}). While
such attacks are important to defend against, this paper focuses on
studying the notion of
perceptual similarity that is relied upon in the majority of the literature on
adversarial examples.}

\section{Necessity and Sufficiency of Conditions for Perceptual Similarity}
\label{sec:limit1}


To effectively find adversarial examples and defend against them,
the parameter choice in \eqnref{eqn:adv_examples} should
help us capture the set of all interesting adversarial examples. In
particular, the selection of $D$ and $\epsilon$ should capture
the samples that are perceptually similar to benign samples.
Ideally, we should be able to define \emph{necessary}
and \emph{sufficient} conditions for perceptual similarity via
$D$ and $\epsilon$.

\subsection{Necessity}

The condition $C:=D(x,\hat{x})\leq \epsilon$ is a necessary condition for
perceptual similarity if:
\[
\hat{x}\ \text{is perceptually similar to}\ x \Rightarrow C
\]

Finding necessary conditions for perceptual similarity is important
for the development of better attacks that find adversarial examples, as
well as better defenses. If the condition $C$ used in hopes of ensuring
perceptual similarity is unnecessary (i.e., there exist examples that are
perceptually similar to benign samples, but do not satisfy $C$),  then the search
space of attacks may be too constrained and some stealthy adversarial examples
may not be found. For defenses, and especially for
adversarial training (because the \dnn{}s are specifically trained to prevent
adversarial inputs that satisfy $C$), unnecessary conditions for perceptual
similarity may lead us to fail at defending against adversarial examples that
do not satisfy $C$.

\begin{figure}
\centering
  \begin{subfigure}{0.23\textwidth}
    \centering
    \includegraphics[width=40px,height=40px]{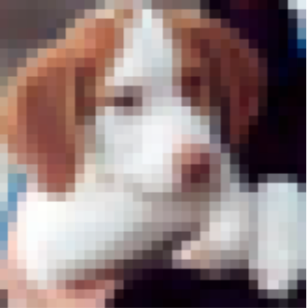}
    \includegraphics[width=40px,height=40px]{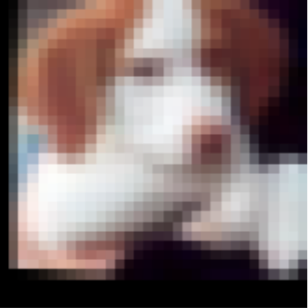}
    \caption{\label{fig:dog}}
  \end{subfigure}
  \begin{subfigure}{0.23\textwidth}
    \centering
    \includegraphics[width=40px,height=40px]{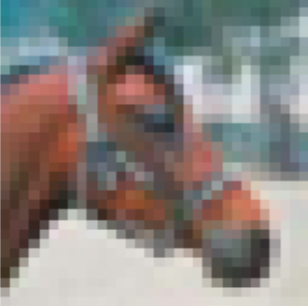}
    \includegraphics[width=40px,height=40px]{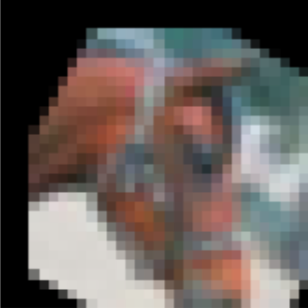}
    \caption{\label{fig:horse}}
  \end{subfigure}
  \caption{Translations and rotations can fool \dnn{}s.
	(a) A dog image (right) resulting from
	transforming a benign image (left) is classified as a cat. (b)
	A horse image (right) resulting from transforming a benign image
	(left) is classified as a truck. Images from Engstrom et
        al.~\cite{Engstrom17AdvTrans}.}
  \label{fig:unnecessary}
\end{figure}

Attacks that craft adversarial examples via applying slight geometric
transformations (e.g., translations and rotations) to benign
samples~\cite{Engstrom17AdvTrans,Kanbak17Transform,Xiao17Transform}
evidence that when using \lpnorm{p}-distance metrics as the measure of
distance, $D$, we may wind up with unnecessary conditions
for perceptual similarity. Such geometric transformations result in
small perceptual differences with respect to benign samples, yet they
result in large \lpnorm{p} distances. \figref{fig:unnecessary}
shows two adversarial examples resulting from
geometric transformation of $32\times 32$ images. While the adversarial
examples are similar to the benign samples, their \lpnorm{p} distance is
large: \lpnorm{0}$\ge$3,010 (maximum possible is 3,072), 
\lpnorm{2}$\ge$15.83 (maximum possible is 55.43), and
\lpnorm{\infty}$\ge$0.87 (maximum possible is 1).\footnote{We use
RGB pixel values in the range [0,1].}
These distances are much larger that what has been used in prior
work (e.g.,~\cite{Carlini17Robustness}).
Indeed, because small \lpnorm{p} is not a necessary condition
for perceptual similarity, state-of-the-art defenses meant to defend
against \lpnorm{p}-bounded adversaries fail at defending against adversarial
examples resulting from geometric transformations~\cite{Engstrom17AdvTrans}.

\subsection{Sufficiency}

The condition $C:=D(x,\hat{x})\leq \epsilon$ is a sufficient condition for
perceptual similarity if:
\[
C \Rightarrow  \hat{x}\ \text{is perceptually similar to}\ x
\]

Alternately, $C$ is insufficient, if it is possible to demonstrate that
a sample $\hat{x}$ is close to $x$ under $D$, while in fact $\hat{x}$
is not perceptually similar to $x$.
The sufficiency of $C$ for perceptual similarity is also important for
both attacks and defenses. In the case of attacks, an adversary using
insufficient conditions for perceptual similarity may craft misclassified
samples that she may deem as perceptually similar to benign samples
under $C$, when they are not truly so. For defenses, the defender may
be attempting to solve an extremely difficult problem by requiring that a
machine-learning model would classify inputs in a certain way, while
even humans may be misled by such inputs due to their lack of
perceptual similarity to benign inputs. In the case of adversarial training,
if $C$ is insufficient, we may even train \dnn{}s to classify inputs
differently than how humans would (thus, potentially poisoning
the \dnn{}). 

In the next section we show that the \lpnorm{p}-norms commonly used
in prior work may be insufficient for ensuring perceptual similarity.
\hlnew{We emphasize that our findings should \emph{not} be interpreted
as stating that the adversarial examples reported on by prior work are not
imperceptible.} 
\hlnew{Instead, our findings highlight that commonly used}
\lpnorm{p}-norms \hlnew{and associated thresholds in principle permit algorithms for crafting
adversarial examples to craft samples that are not perceptually similar
to benign ones, leading to the
undesirable outcomes described above. Specific instances of algorithms
that use those} \lpnorm{p}-norms \hlnew{and thresholds could still
produce imperceptible adversarial examples because the creation of
these examples is constrained by factors other than the chosen}
\lpnorm{p}-norms and thresholds.

\section{Experiment Design and Results }
\label{sec:studies}


To show that a small \lpnorm{p} distance ($p\in \{0, 2, \infty\}$) from benign
samples is insufficient for ensuring perceptual similarity to these samples,
we conducted three online user studies (one for each $p$). The goal of each study
was to show that, for small values of $\epsilon$, it is
possible to find samples that are close to benign samples in \lpnorm{p}, but
are not perceptually similar to those samples to a human. In what follows, 
we present the high-level experimental design that is common among the
three studies. Next, we report on our study participants. Then, we provide the
specific design details for each study and the results.

\subsection{Experiment Design}

Due to the many ways in which two images can be similar, it is unclear
whether one can learn useful input from users by directly asking them
about the level of similarity between image pairs. Therefore, we rely on
indirect reasoning to determine whether an image $\hat{x}$ is perceptually
similar to $x$. In particular, we make the following observation: if we ask
mutually exclusive sets of users about the contents of $\hat{x}$ and $x$,
and they disagree, then we learn that $x$ and $\hat{x}$ are \emph{definitely}
dissimilar; otherwise, we learn that they are \emph{likely} similar. Our observation
is motivated by Papernot et al.'s approach to determine that their attack is
imperceptible to humans~\cite{Papernot16Limitations}.

Motivated by the above-mentioned observation, we followed a
between-subject design for each study, assigning each participant
to one of three conditions: \benign{}, \advimp{}, and \advper{}. In all the
conditions, participants were shown images and were asked to select
the label (i.e., category) that best describes the image out of ten possible
labels (e.g., the digit shown in the image). The conditions differed in
the nature of images shown to the participants. Participants in
\benign{} (``benign'' condition) were shown unaltered images from
standard datasets. Participants in \advimp{} (``adversarial and
imperceptible'' condition) were shown adversarial examples of
images in \benign{} that fool state-of-the-art \dnn{}s. The images
in \advimp{} have small \lpnorm{p} distances to images in \benign{},
and \emph{were not} designed to mislead humans. Participants in 
\advper{} (``adversarial and perceptible'' condition) were shown
variants of the images in \benign{} that are close in \lpnorm{p} distance
to their counterparts in \benign{}, but were designed to
mislead both humans and \dnn{}s. To lower the mental
burden on participants, each participant was asked only 25
image-categorization questions. Because the datasets
we used contain thumbnail images (see below), we presented the
images to participants in three different sizes: original size, resized
$\times 2$, and resized $\times 4$. Additionally to categorizing
images, we asked participants in all conditions about
their level of confidence in their answers on a 5-point Likert
scale (one denotes low confidence and five denotes high confidence).
\hlnew{The protocol was approved by Carnegie Mellon's review board.}

Conceptually, the responses of participants in \benign{} help
us estimate humans' accuracy on benign images (i.e., their likelihood to
pick the labels consistent with the ground truth). By comparing
the accuracy of users in \advimp{} to \benign{}, we learn whether
the attack and the threshold on \lpnorm{p} distance that we pick
result in imperceptible attacks. We hypothesize that images in
\advimp{} are likely to be categorized correctly by users. Hence,
the attack truly crafts adversarial examples, and poses risk to the
integrity of \dnn{}s at the chosen threshold. In contrast, by comparing
\advper{} with \benign{}, we hope to show that, for the same threshold,
it is possible to find instances that mislead humans and \dnn{}s. Namely,
we hypothesize that the accuracy of users on \advper{} is significantly
lower than their accuracy on \benign{}. If our hypothesis is validated, we
learn that small \lpnorm{p} distance does not ensure perceptual similarity.

\parheading{Datasets and \dnn{}s.}
In the studies, we used images from the MNIST~\cite{MNIST}
and CIFAR10~\cite{Krizhevsky09CIFAR} datasets. MNIST is a
dataset of $28\times 28$ pixel images of digits, while CIFAR10 is
a dataset of $32\times 32$ pixel images that contain ten object
categories: airplanes, automobiles, birds,  cats, deer, dogs,
frogs, horses, ships, and trucks. Both MNIST and CIFAR10
are widely used for developing both attacks on \dnn{}s and
defenses against them (e.g.,~\cite{Gdfllw14ExpAdv,
Madry17AdvTraining,Papernot16Limitations}).

In conditions \advimp{} and \advper{}, we created attacks against
two \dnn{}s published by Madry et
al.~\cite{Madry17AdvTraining}---one for MNIST, and another
for CIFAR10. The MNIST \dnn{} is highly accurate, achieving 98.8\%
accuracy on the MNIST test set. The CIFAR10 \dnn{} also achieves
a relatively high accuracy---87.3\% on CIFAR10's test
set. More notably, both \dnn{}s are two of the most resilient models
to adversarial examples (specifically, ones with bounded
\lpnorm{\infty} distance from benign samples) known to date.

\subsection{Participants} 

We recruited a total of 399 participants from
the United States through the Prolific crowdsourcing platform.
Their ages
ranged from 18 to 78, with a mean of 32.85 and standard deviation of 11.54.
The demographics of our participants were slightly skewed toward males:
59\% reported to be males, 39\% reported to be females, and 1\% preferred
not to specify their gender. 34\% of the participants were students, and
84\% were employed at least partially. \hlnew{Participants took an average
of roughly six minutes to complete the study and were compensated \$1.}

\begin{figure}
\centering
	\newcommand\studyfigwidth[0]{0.32\textwidth}
	\begin{subfigure}[t]{0.15\textwidth}
	\begin{tiny}
	\centering
	\begin{tabu} to 1 \textwidth {X[1,c] X[1,c] X[1,c]}
	\benign{} & \advimp{} & \advper{}\\[0.1cm]
	\includegraphics[width=\studyfigwidth]{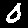}
		 & \includegraphics[width=\studyfigwidth]{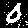}
		 & \includegraphics[width=\studyfigwidth]{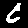}\\
	\includegraphics[width=\studyfigwidth]{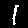}
		 & \includegraphics[width=\studyfigwidth]{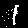}
		 & \includegraphics[width=\studyfigwidth]{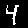}\\
	\includegraphics[width=\studyfigwidth]{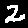}
		 & \includegraphics[width=\studyfigwidth]{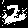}
		 & \includegraphics[width=\studyfigwidth]{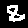}\\
	\includegraphics[width=\studyfigwidth]{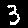}
		 & \includegraphics[width=\studyfigwidth]{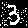}
		 & \includegraphics[width=\studyfigwidth]{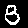}\\
	\includegraphics[width=\studyfigwidth]{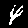}
		 & \includegraphics[width=\studyfigwidth]{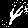}
		 & \includegraphics[width=\studyfigwidth]{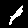}\\
	\includegraphics[width=\studyfigwidth]{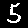}
		 & \includegraphics[width=\studyfigwidth]{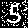}
		 & \includegraphics[width=\studyfigwidth]{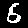}\\
	\includegraphics[width=\studyfigwidth]{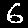}
		 & \includegraphics[width=\studyfigwidth]{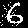}
		 & \includegraphics[width=\studyfigwidth]{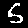}\\
	\includegraphics[width=\studyfigwidth]{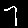}
		 & \includegraphics[width=\studyfigwidth]{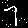}
		 & \includegraphics[width=\studyfigwidth]{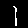}\\
	\includegraphics[width=\studyfigwidth]{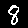}
		 & \includegraphics[width=\studyfigwidth]{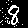}
		 & \includegraphics[width=\studyfigwidth]{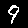}\\
	\includegraphics[width=\studyfigwidth]{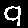}
		 & \includegraphics[width=\studyfigwidth]{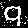}
		 & \includegraphics[width=\studyfigwidth]{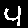}\\
	\end{tabu}
	\caption{\label{fig:l0_samples}\lpnorm{0}}
	\end{tiny}
	\end{subfigure}
	\begin{subfigure}[t]{0.15\textwidth}
	\begin{tiny}
	\centering
	\begin{tabu} to 1 \textwidth {X[1,c] X[1,c] X[1,c]}
	\benign{} & \advimp{} & \advper{}\\[0.1cm]
	\includegraphics[width=\studyfigwidth]{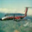}
		 & \includegraphics[width=\studyfigwidth]{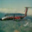}
		 & \includegraphics[width=\studyfigwidth]{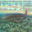}\\
	\includegraphics[width=\studyfigwidth]{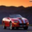}
		 & \includegraphics[width=\studyfigwidth]{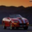}
		 & \includegraphics[width=\studyfigwidth]{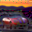}\\
	\includegraphics[width=\studyfigwidth]{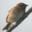}
		 & \includegraphics[width=\studyfigwidth]{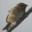}
		 & \includegraphics[width=\studyfigwidth]{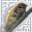}\\
	\includegraphics[width=\studyfigwidth]{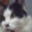}
		 & \includegraphics[width=\studyfigwidth]{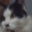}
		 & \includegraphics[width=\studyfigwidth]{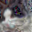}\\
	\includegraphics[width=\studyfigwidth]{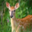}
		 & \includegraphics[width=\studyfigwidth]{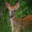}
		 & \includegraphics[width=\studyfigwidth]{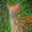}\\
	\includegraphics[width=\studyfigwidth]{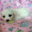}
		 & \includegraphics[width=\studyfigwidth]{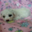}
		 & \includegraphics[width=\studyfigwidth]{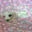}\\
	\includegraphics[width=\studyfigwidth]{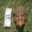}
		 & \includegraphics[width=\studyfigwidth]{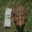}
		 & \includegraphics[width=\studyfigwidth]{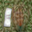}\\
	\includegraphics[width=\studyfigwidth]{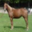}
		 & \includegraphics[width=\studyfigwidth]{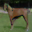}
		 & \includegraphics[width=\studyfigwidth]{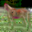}\\
	\includegraphics[width=\studyfigwidth]{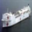}
		 & \includegraphics[width=\studyfigwidth]{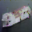}
		 & \includegraphics[width=\studyfigwidth]{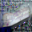}\\
	\includegraphics[width=\studyfigwidth]{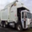}
		 & \includegraphics[width=\studyfigwidth]{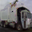}
		 & \includegraphics[width=\studyfigwidth]{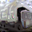}\\
		
	\end{tabu}
	\caption{\label{fig:l2_samples}\lpnorm{2}}
	\end{tiny}
	\end{subfigure}
	\begin{subfigure}[t]{0.15\textwidth}
	\begin{tiny}
	\centering
	\begin{tabu} to 1 \textwidth {X[1,c] X[1,c] X[1,c]}
	\benign{} & \advimp{} & \advper{}\\[0.1cm]
	\includegraphics[width=\studyfigwidth]{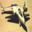}
		 & \includegraphics[width=\studyfigwidth]{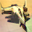}
		 & \includegraphics[width=\studyfigwidth]{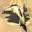}\\
	\includegraphics[width=\studyfigwidth]{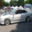}
		 & \includegraphics[width=\studyfigwidth]{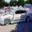}
		 & \includegraphics[width=\studyfigwidth]{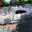}\\
	\includegraphics[width=\studyfigwidth]{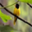}
		 & \includegraphics[width=\studyfigwidth]{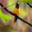}
		 & \includegraphics[width=\studyfigwidth]{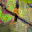}\\
	\includegraphics[width=\studyfigwidth]{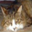}
		 & \includegraphics[width=\studyfigwidth]{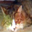}
		 & \includegraphics[width=\studyfigwidth]{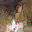}\\
	\includegraphics[width=\studyfigwidth]{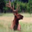}
		 & \includegraphics[width=\studyfigwidth]{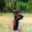}
		 & \includegraphics[width=\studyfigwidth]{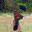}\\
	\includegraphics[width=\studyfigwidth]{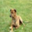}
		 & \includegraphics[width=\studyfigwidth]{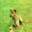}
		 & \includegraphics[width=\studyfigwidth]{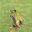}\\
	\includegraphics[width=\studyfigwidth]{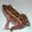}
		 & \includegraphics[width=\studyfigwidth]{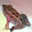}
		 & \includegraphics[width=\studyfigwidth]{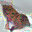}\\
	\includegraphics[width=\studyfigwidth]{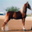}
		 & \includegraphics[width=\studyfigwidth]{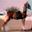}
		 & \includegraphics[width=\studyfigwidth]{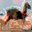}\\
	\includegraphics[width=\studyfigwidth]{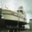}
		 & \includegraphics[width=\studyfigwidth]{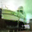}
		 & \includegraphics[width=\studyfigwidth]{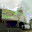}\\
	\includegraphics[width=\studyfigwidth]{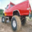}
		 & \includegraphics[width=\studyfigwidth]{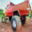}
		 & \includegraphics[width=\studyfigwidth]{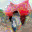}\\

	\end{tabu}
	\caption{\label{fig:linf_samples}\lpnorm{\infty}}
	\end{tiny}
	\end{subfigure}
	\caption{\label{fig:study_samples}Sample images from the three conditions 
	we had for each \lpnorm{p}-norm. Each row shows three variants of the 
	same image. 
	}
\end{figure}

\begin{figure*}
\centering
  \begin{subfigure}[t]{0.32\textwidth}
    \centering
    \includegraphics[width=\textwidth]{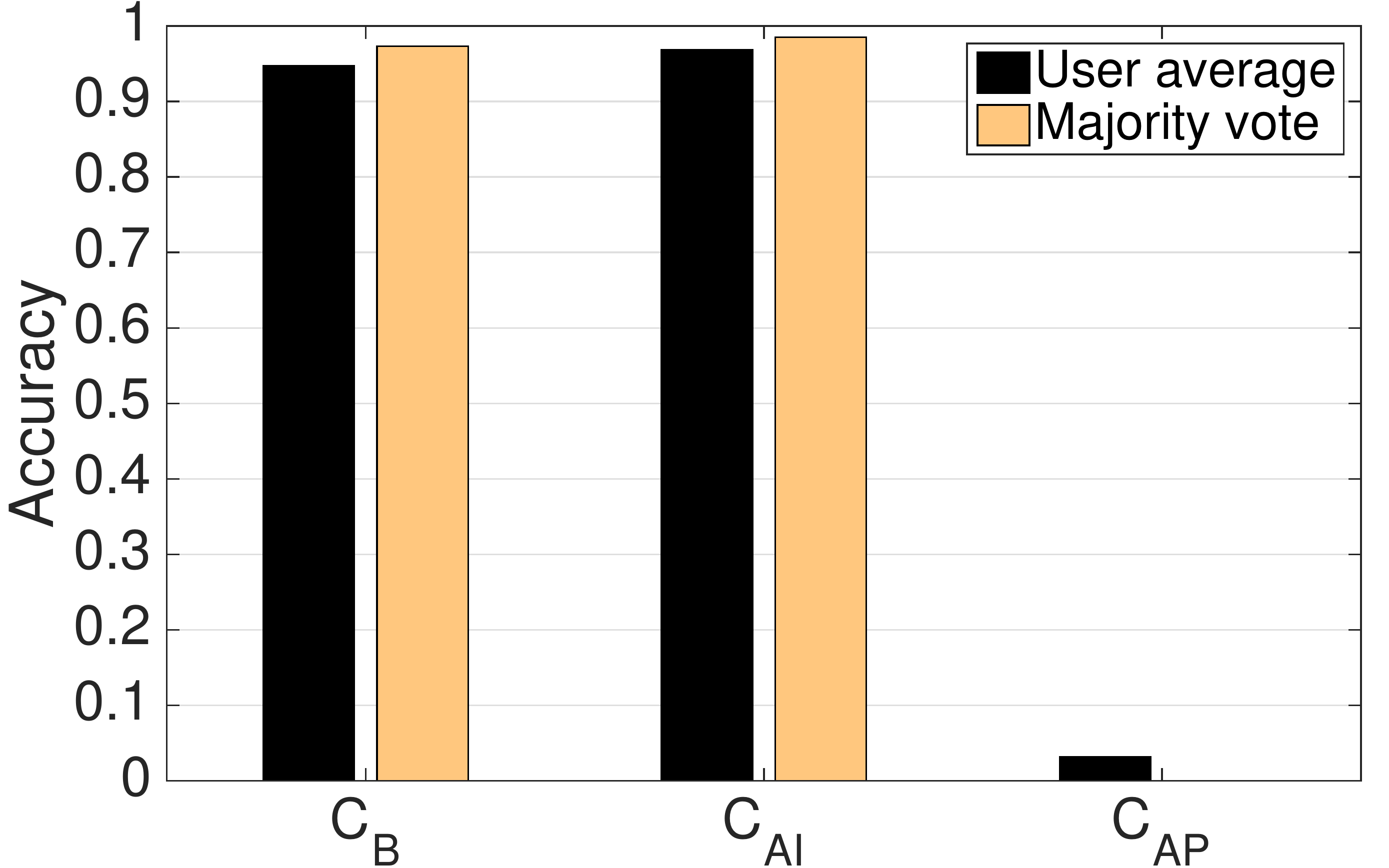}
    \caption{\label{fig:l0_res}\lpnorm{0}}
  \end{subfigure}
  \begin{subfigure}[t]{0.32\textwidth}
    \centering
    \includegraphics[width=\textwidth]{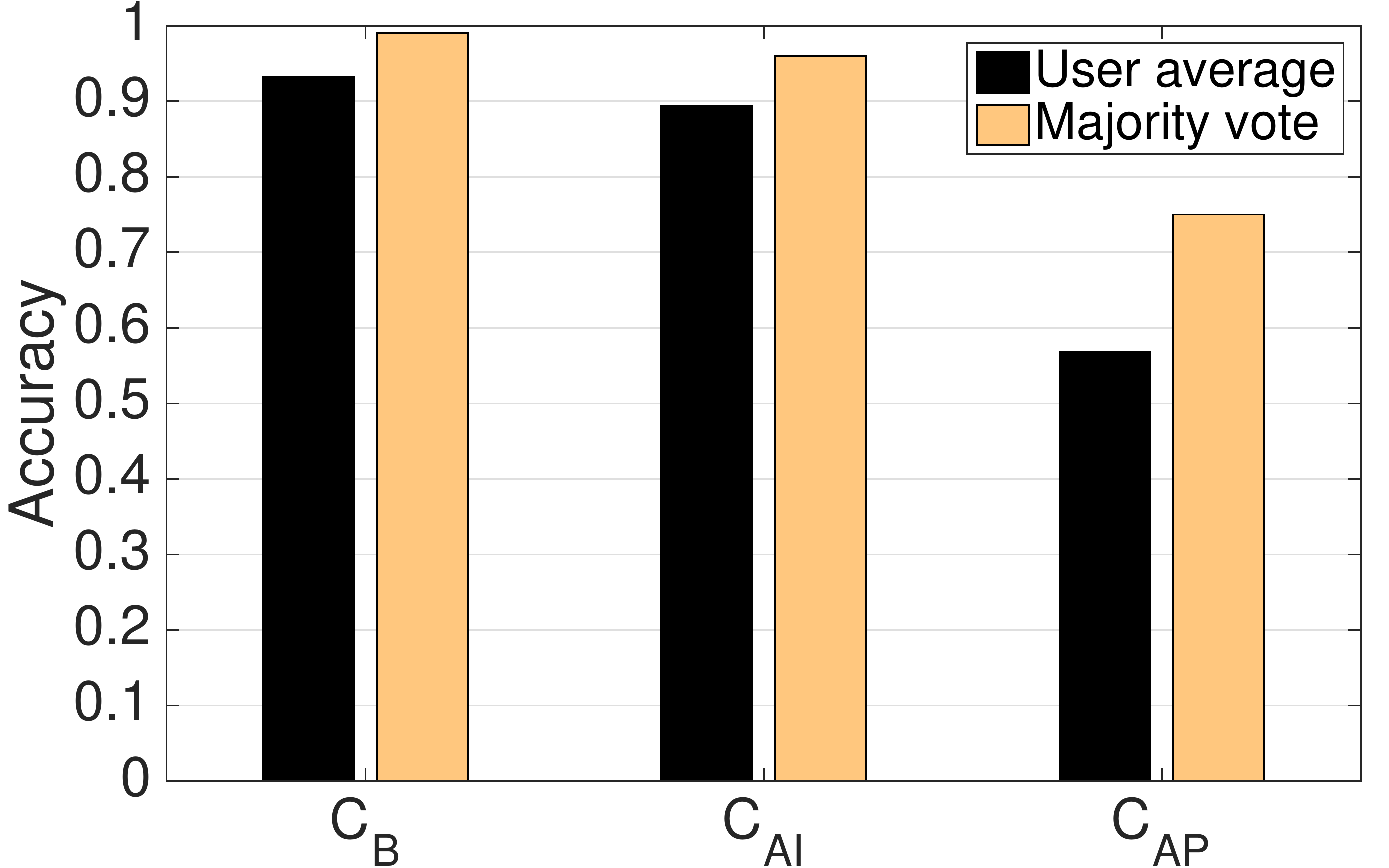}
    \caption{\label{fig:l2_res}\lpnorm{2}}
  \end{subfigure}
  \begin{subfigure}[t]{0.32\textwidth}
    \centering
    \includegraphics[width=\textwidth]{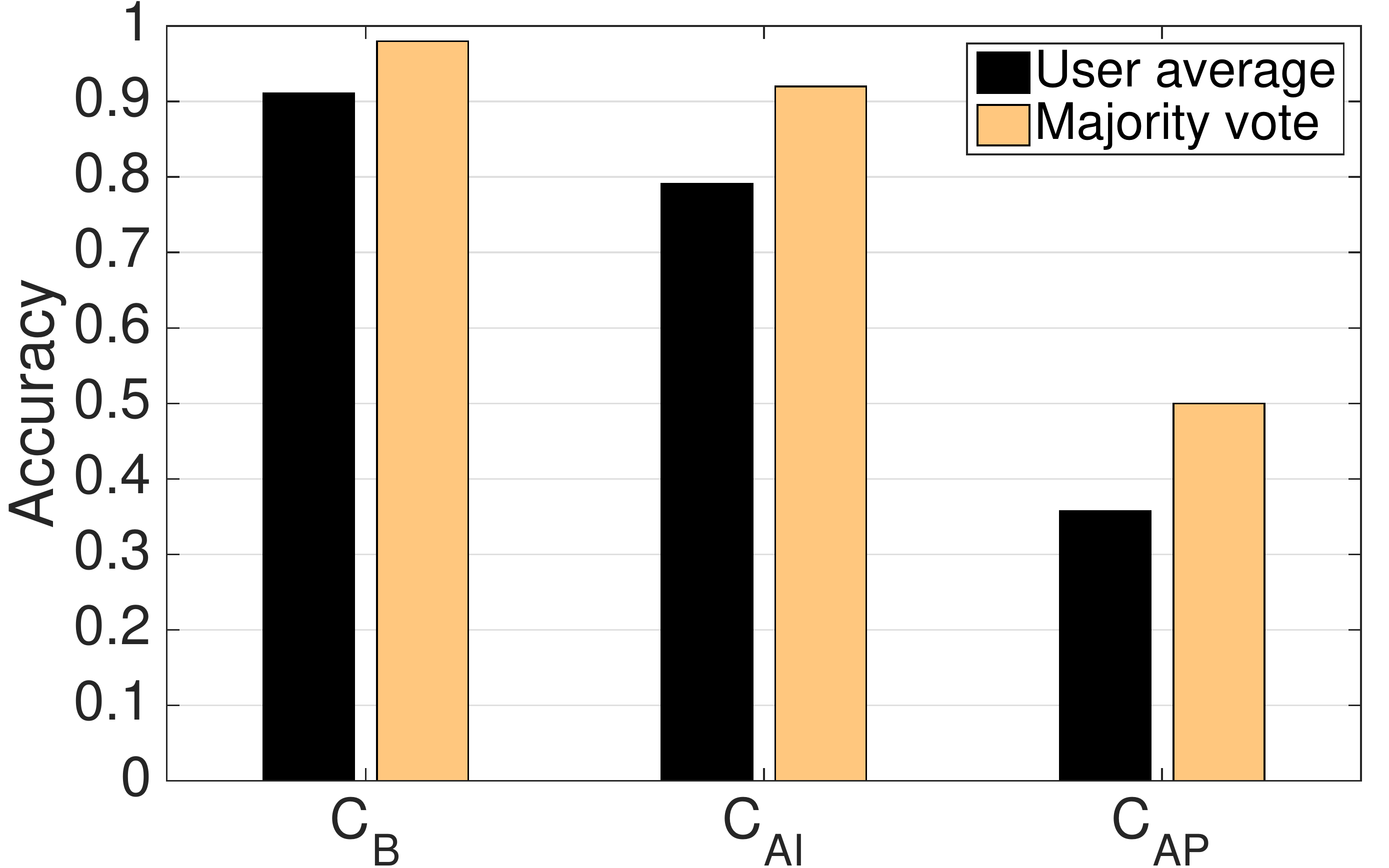}
    \caption{\label{fig:linf_res}\lpnorm{\infty}}
  \end{subfigure}
  \caption{User performance for the three \lpnorm{p}-norms that we studied.
	For each condition (\benign{}, \advimp{}, and \advper{}),  we report the users'
	average accuracy, and the accuracy when labeling each image by the majority
	vote (over the labels provided by the participants). Accuracy is the fraction
	of labels that are consistent with the ground truth.}
  \label{fig:results}
\end{figure*}

\subsection{Insufficiency of \lpnorm{0}}

\parheading{Study Details.}
We used the MNIST dataset and \dnn{} to test whether it is
possible to perturb images only slightly on \lpnorm{0} while
simultaneously misleading humans and \dnn{}s.
\benign{} was assigned 75 randomly selected images from the test
set of MNIST. All 75 images were correctly classified by the \dnn{}.
We used the Jacobian Saliency Map Approach
(JSMA)~\cite{Papernot16Limitations}, as implemented in
Cleverhans~\cite{Papernot17Cleverhans}, to craft adversarial
examples for \advimp{}. We limited the amount of change applied to
an image to at most 15\% of the pixels, and found that JSMA was able
to find successful attacks for 68 out of 75 images. For successful
attacks, JSMA perturbed 4.90\% (2.91\% standard deviation) of pixels
on average---this is comparable to the result of Papernot et
al.~\cite{Papernot16Limitations}. Images in \advper{} were
crafted manually. Using a photo-editing software\footnote{GIMP
(\url{https://www.gimp.org/})} while simultaneously
receiving feedback from the DNN, a member of our team edited the
75 images from \benign{} while
attempting to minimize the number of pixels changed
such that the resulting image would be misclassified by both
humans and the \dnn{}. The average \lpnorm{0} distance of
images in \advper{} from their counterparts in \benign{} is 4.48\%
(2.45\% standard deviation). Examples of the images in the
three conditions are shown in \figref{fig:l0_samples}.
We note that because creating the images for \advper{} involves
time-consuming manual effort, we limited each condition to at
most 75 images.

A total of 201 participants were assigned to the \lpnorm{0} study:
73 were assigned to \benign{}, 59 to \advimp{}, and 69 to \advper{}.

\newpage

\parheading{Experiment results.}
We computed the users' accuracy
(i.e., how often their responses agreed with the ground truth), and the
accuracy when classifying each image according to the majority vote
over all labels provided by the users. \figref{fig:l0_res} shows the
results. We found that users' average accuracy was high for the
unaltered images of \benign{} (95\%) and adversarial images of
\advimp{} (97\%), but low for the images of \advper{} (3\%). In fact,
if we classify images via majority votes, none of the images of \advper{}
would be classified correctly. The difference between users' average
accuracy in \benign{} and \advimp{} is not statistically significant
according to a t-test ($p=0.34$). In contrast, the difference between
users' average accuracy in \advper{} and other conditions is
statistically significant ($p<0.01$).  Users in all the conditions were
confident in their responses---the average confidence levels ranged
from 4.19 (\advper{}) to 4.47 (\benign{}).

The results support our hypotheses. While it is possible
to find adversarial examples with small \lpnorm{0} distance from benign
samples, it is possible, for the same distance, to find samples that are
not imperceptible to humans. In fact, humans may be highly
confident that those samples belong to other classes.

\subsection{Insufficiency of \lpnorm{2}}

\parheading{Study Details.}
We used the CIFAR10 dataset and \dnn{} to test \lpnorm{2} for
insufficiency. We randomly picked 100 images from the test set
that were correctly classified by the \dnn{} for \benign{}.
For each image in \benign{}, we created (what we hoped would
be) an imperceptible adversarial example for \advimp{}. Images
in \advimp{} have a fixed \lpnorm{2} distance of 6 from their
counterparts in \benign{}. Because we did not find evidence in
the literature for an upper bound on \lpnorm{2} distance
that is is still imperceptible to humans, we chose a distance of 6
empirically---our results (below) support this choice. To create the
adversarial examples, we used an iterative gradient descent approach,
in the vein of prior work~\cite{Carlini17Robustness}, albeit
with two notable differences. First, we used an algorithm by
Wang et al.~\cite{Wang04SSIM} to initialize the attack to an
image that has high SSIM to the benign image, but lies at a fixed
\lpnorm{2} distance from it. The rationale
behind this was to increase the perceptual similarity
between the adversarial image and the benign image.
Second, we ensured that the \lpnorm{2}-norm of the perturbation
is fixed by normalizing it to 6 after each iteration of the attack.
The images of \advper{} we generated via a similar approach
to those of \advimp{}. The only difference is that we initialized
the attack with an image that has low SSIM with respect to its
counterpart benign image using Wang et al.'s algorithm.
\figref{fig:l2_samples} shows a sample of the images
that we used in the \lpnorm{2} study.

In total, we had 99 participants assigned to the \lpnorm{2} study:
25 were assigned to \benign{}, 38 to \advimp{}, and 36 to \advper{}.

\parheading{Experiment results.}
Users' average accuracy and the accuracy of the majority
vote are shown in \figref{fig:l2_res}. On the benign images
of \benign{}, users had an average accuracy of 93\%. Their
average accuracy on the images of \advimp{} was 89\%, not
significantly lower ($p\approx 1$). Moreover, we found that 
users in \benign{} and \advimp{} were almost equally confident
about their choices (averages of 4.31 and 4.37).
We thus concluded that 6 is a reasonable bound for \lpnorm{2}
attacks. In stark contrast, we found that \advper{}
users' average accuracy dropped to 57\% and confidence to
2.97 ($p<0.01$ for both). In
other words, users' likelihood to make mistakes increased
by 36\%, on average, and their confidence in their decisions
decreased remarkably.

The results support our hypotheses, as a significant fraction of
attack samples with bounded \lpnorm{2} can be perceptually
different than their corresponding benign samples.

\subsection{Insufficiency of \lpnorm{\infty}}

\parheading{Study Details.}
Similarly to the \lpnorm{2} study, we used the CIFAR10
dataset and \dnn{} also for the \lpnorm{\infty}
study. We again picked 100 random images from the test set for
\benign{}. For \advimp{}, we generated adversarial examples
with \lpnorm{\infty} distance of 0.1 from benign examples, as
done by Goodfellow et al.~\cite{Gdfllw14ExpAdv}. We generated
the adversarial examples using the Projected Gradient Descent
algorithm~\cite{Madry17AdvTraining}, with a simple tweak to
enhance imperceptibility: after each gradient descent
iteration, we increased SSIM with respect to benign images
using Wang et al.'s algorithm~\cite{Wang04SSIM}. 
Examples in \advper{} were generated using a similar
algorithm, but we decreased the SSIM with respect to benign
images instead of increasing it.

In total, we had 99 participants assigned to the study:
36 were assigned to \benign{}, 31 to \advimp{}, and 32 to
\advper{}.

\parheading{Experiment results.}
The accuracy results are summarized in \figref{fig:linf_res}.
On benign examples, users' average accuracy was 91\%.
Their average confidence score was 4.45. The attacks in
\advimp{} were not completely imperceptible:
users' average accuracy decreased to 79\% and confidence score
to 3.98 (both significant with $p<0.01$).
However, attacks in \advper{} were significantly less similar
to benign images: users' average accuracy and confidence
score were 36\% and 3.04 ($p<0.01$).

Our hypotheses hold also for \lpnorm{\infty}---a significant
fraction of attacks with relatively small \lpnorm{\infty} can be
perceptually different to humans than benign images.

\section{Discussion}
\label{sec:discussion}

The results of the user studies confirm our hypotheses---defining
similarity using \lpnorm{0}, \lpnorm{2}, and \lpnorm{\infty}-norms
can be insufficient for ensuring perceptual similarity in some cases.
Here, we discuss some of the limitations of this work, and discuss
some alternatives for \lpnorm{p}-norms.

\subsection{Limitations}

A couple of limitations should be considered when interpreting our results.
First, we demonstrated our results on two \dnn{}s and for two datasets,
and so they may not apply for every \dnn{} and image-recognition task.
The \dnn{}s that we considered are among the most resilient models
to adversarial examples to date. Consequently, we believe that
the attacks we generated against them for \advimp{} and \advper{}
would succeed against other \dnn{}s. 
\hlnew{Depending on the chosen thresholds,
our results may or may not directly apply to specific combinations
of norms and
image-recognition tasks that we did not consider in this work.
While studying more combinations may be useful, we believe that our
findings are impactful in their current form, as the combinations
of norms, thresholds, and datasets we considered are commonly used in practice
(e.g.,~\cite{Carlini17Robustness,Gdfllw14ExpAdv,
Papernot16Limitations,Szegedy13NNsProps}).}
\hlnew{We note that it may be possible to achieve sufficient conditions
for perceptual similarity by using lower thresholds than in our experiments.
Stated differently, it may be impossible to fool humans using lower thresholds.
However, decreasing thresholds may also prevent algorithms for crafting attacks
from finding successful adversarial examples (namely, the ones that are 
part of \advimp{} in the experimental conditions).}

Second, we estimated similarity using a proxy: whether participants'
categorizations of perturbed images were consistent with their categorizations
of their benign counterparts. However, similarity has different facets that 
may or may not be interesting, depending on the threat model being
studied. For instance, in some cases we may want to estimate whether
certain attacks are inconspicuous or not (e.g., to learn whether TSA
agents would detect disguised individuals attempting to mislead
surveillance systems). In such cases, we want to measure whether
adversarial images are ``close enough'' to benign images to the extent
that a human observer cannot reliably tell between adversarial and
non-adversarial images. We believe that future work should develop
a better understanding of this and other notions of similarity and how
to assess them as a means to help us improve current attacks and
defenses.

\subsection{Alternatives to \lpnorm{p}-norms}

\hlnew{We next list several alternative distance metrics for assessing
  similarity and provide a preliminary assessment of their suitability
  as a replacement for \lpnorm{p}-norms.}

Our results show that by using the same threshold for all samples,
one may generate adversarial images that mislead humans---thus,
they are not imperceptible. As a solution, one may consider
setting a different threshold for every sample to ensure that the
attacks' output would be imperceptible. In this case, a
principled automated method would be needed to set the
sample-dependent thresholds in order to create attacks at scale.
Without such a method, human feedback may be needed in the
process for every sample.

\begin{figure}[t]
\centering
    \begin{subfigure}[t]{0.12\textwidth}
    \centering
    \includegraphics[width=\textwidth]{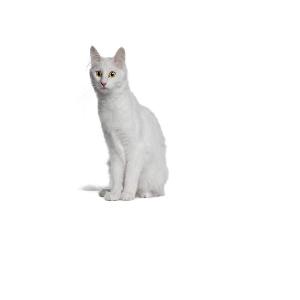}
    \includegraphics[width=\textwidth]{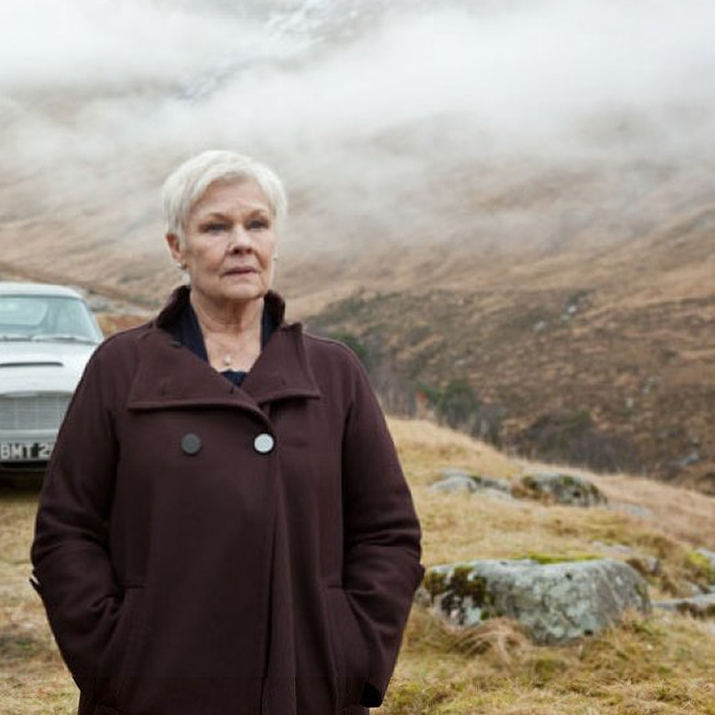}
    \caption{\label{fig:cat_ssim}}
  \end{subfigure}
  \begin{subfigure}[t]{0.12\textwidth}
    \centering
    \includegraphics[width=\textwidth]{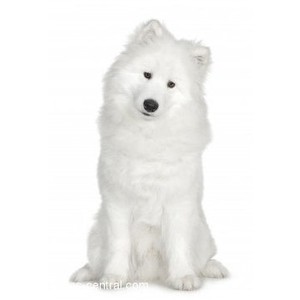}
    \includegraphics[width=\textwidth]{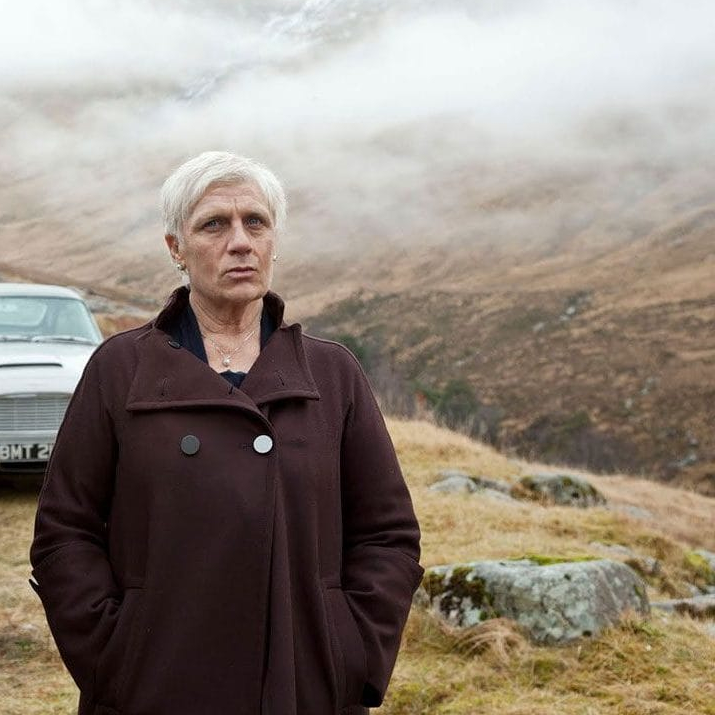}
    \caption{\label{fig:dog_ssim}}
  \end{subfigure}
  \begin{subfigure}[t]{0.12\textwidth}
    \centering
    \includegraphics[width=\textwidth]{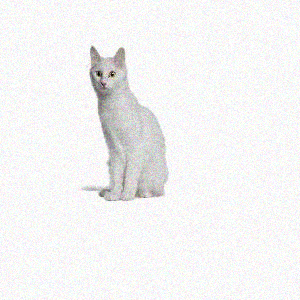}
    \includegraphics[width=\textwidth]{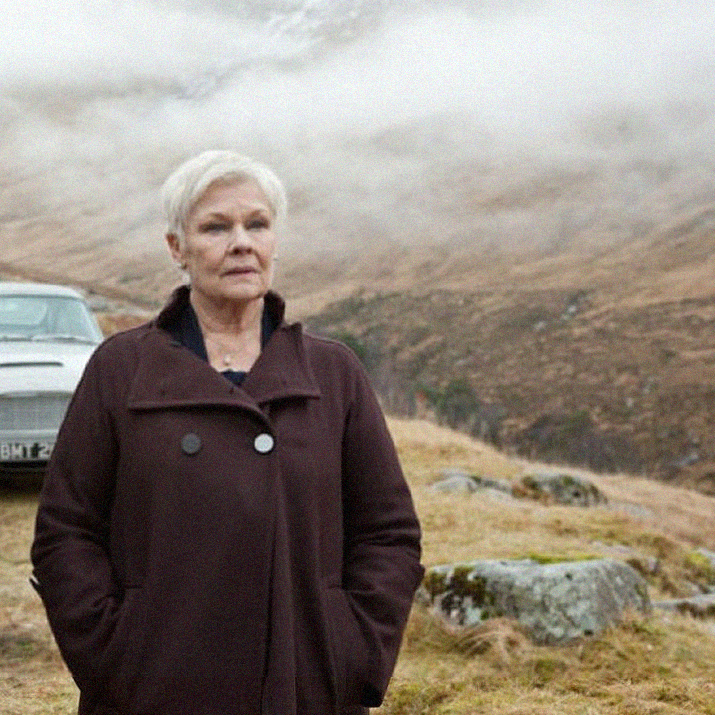}
    \caption{\label{fig:cat_r_ssim}}
  \end{subfigure}
\caption{\label{fig:bad_ssim} SSIM can be high between two
	images containing different objects or subjects. Despite
	showing different animals or subjects (the bottom image
	in (b) was created by swapping the faces of actress Judi Dench
	and actor Daniel Craig), the SSIM value between the images
	in (a) and (b) is high---0.89 between the top images, and 0.95
	between the bottom images. Images in (c) were created by
	adding uniform noise to images in (a).
	The SSIM value between (a) and (c) is 0.77 for the images
	at the top, and 0.87 for the images at the bottom. (Sources
	of images in (a) and (b): \url{https://goo.gl/Mxo9mK},
	\url{https://goo.gl/GEd6Bs},  \url{https://goo.gl/mvvFZ1},
	and \url{https://goo.gl/vwuK9t}.)}
\end{figure}

Other similarity-assessment measures, such as SSIM~\cite{Wang04SSIM}
and the ``minimal'' transform needed to align two images, that have been
used in prior work (e.g.,~\cite{Kanbak17Transform,Rozsa16SSIM}) may be considered
to replace \lpnorm{p}-norms. Additionally, one may consider image-similarity
metrics that have not been previously used in the adversarial examples
literature, such as Perceptual Image Diff~\cite{Yee0PerDiff} and the
Universal Quality Index~\cite{Wang02ImageQuality}.  Such metrics should
be treated with care, as they may lead to unnecessary and insufficient conditions for
perceptual similarity. For example, SSIM is sensitive to small geometric transformation
(e.g., the SSIM between the images in \figref{fig:horse} is 0.36 out of 1, which is
relatively low~\cite{Wang04SSIM}). So, using SSIM to define similarity
may lead to unnecessary conditions. Moreover, as demonstrated in
\figref{fig:bad_ssim}, SSIM may be high even when two images are not similar.
Thus, SSIM may lead to insufficient conditions for similarity.

\hlnew{The recent work of Jang et al.~suggests three metrics
to evaluate the similarity between adversarial examples
and benign inputs~\cite{Jang17Objective}. The metrics evaluate
adversarial perturbations' quality by how much they corrupt the Fourier
transform, their effect on edges, or their effect on the images' gradients.
This work appears to take a step in the right direction. However, further validation
of the proposed metrics is needed. For instance, we speculate
that slight geometric transformations to a benign image might affect the
gradients in the image dramatically. Thus, metrics based on gradients
alone may be unnecessary for defining conditions for perceptual similarity.}

Lastly, one may consider using several metrics simultaneously to define
similarity (e.g., by allowing geometric transformations and perturbations
with small \lpnorm{p}-norm to craft adversarial
examples~\cite{Engstrom17AdvTrans}). While this may be
a promising direction, metrics should be combined with
special care. As the conjunction of one or more unnecessary conditions leads
to an unnecessary condition, and the disjunction of one or more insufficient
conditions leads to an insufficient condition, simply conjoining or disjoining
metrics may not solve the (in)sufficiency and (un)necessity problems
of prior definitions of similarity.

\hlnew{Finding better measures for assessing perceptual similarity
remains an open problem. Better similarity measures could help improve both 
algorithms for finding adversarial examples and, more importantly,
defenses against them. In the absence of
such measures, we recommend that future research rely not
only on known metrics for perceptual similarity assessment, but also
on human-subject studies.}

\section{Conclusion}
\label{sec:conclusion}

In this work, we aimed to develop a better understanding of
the suitability of \lpnorm{p}-norms for measuring the similarity
between adversarial examples and benign images. Specifically,
we complemented our knowledge that conditions on \lpnorm{p}
distances used for defining similarity are unnecessary in some
cases---i.e., they may not capture all imperceptible
attacks---and showed that they can also be insufficient---i.e.,
they may lead one to conclude that an adversarial instance is
similar to a benign instance when it is not so. Hence,
\lpnorm{p}-distance metrics may be unsuitable for assessing similarity
when crafting adversarial examples and defending against them.
We pointed out possible alternatives for \lpnorm{p}-norms to
assess similarity, though they seem to have limitations, too.
Thus, there is a need for further research to improve the
assessment of similarity when developing attacks and
defenses for adversarial examples.

\section*{Acknowledgments}

This work was supported in part by MURI grant W911NF-17-1-0370, by the
National Security Agency, by a gift from NVIDIA, and gifts from NATO
and Lockheed Martin through Carnegie Mellon CyLab.

{\small
\bibliographystyle{ieee}
\bibliography{cited}
}

\end{document}